\documentclass[twocolumn,showpacs,amstex,amssymb,prb]{revtex4}
\usepackage{graphicx}
\usepackage{epstopdf}
\newcommand{\rr}{{\bf r}}
\newcommand{\e}{{\bf e}}
\begin{document}
\title{The analytic structure of Bloch functions for linear molecular chains}
\author{E. Prodan}
\address{PRISM, Princeton University, Princeton, NJ 08544}
\email{eprodan@princeton.edu}

\begin{abstract}
This paper deals with Hamiltonians of the form $H=-{\bf \nabla}^2+v(\rr)$, with $v(\rr)$ periodic along the $z$ direction, $v(x,y,z+b)=v(x,y,z)$. The wavefunctions of $H$ are the well known Bloch functions $\psi_{n,\lambda}(\rr)$, with the fundamental property $\psi_{n,\lambda}(x,y,z+b)=\lambda \psi_{n,\lambda}(x,y,z)$ and $\partial_z\psi_{n,\lambda}(x,y,z+b)=\lambda 
\partial_z\psi_{n,\lambda}(x,y,z)$. We give the generic analytic structure (i.e. the Riemann surface) of $\psi_{n,\lambda}(\rr)$ and their corresponding energy, $E_n(\lambda)$, as functions of $\lambda$. We show that $E_n(\lambda)$ and $\psi_{n,\lambda}(x,y,z)$ are different branches of two multi-valued analytic functions, $E(\lambda)$ and $\psi_\lambda(x,y,z)$, with an essential singularity at $\lambda=0$ and additional branch points, which are generically of order 1 and 3, respectively. We show where these branch points come from, how they move when we change the potential and how to estimate their location. Based on these results, we give two applications: a compact expression of the Green's function and a discussion of the asymptotic behavior of the density matrix for insulating molecular chains.
\end{abstract}

\pacs{71.10.-w, 71.15.-m}

\maketitle
\section{Introduction}
The analytic structure of the Bloch functions for 1D crystals with inversion symmetry was investigated in Ref.~\onlinecite{Kohn59}. Among the major conclusions of this paper, is the fact that the entire band structure can be characterized by a single, though multi-valued, analytic function $E(\lambda)$[$\lambda=e^{ikb}$], with branch points that occur at complex $k$. A similar conclusion holds also for the Bloch functions. The positions of the branch points determine the exponential decay of the Wannier functions. For insulators, they also determine the exponential decay of the density matrix and other correlation functions. It was recently shown that the order of the branch points determines the additional inverse power law decay of these functions.\cite{Vanderbilt01} We can say that, although the branch points occur at complex $k$, their existence and location have very important implications on the properties and dynamics of the physical states. The complex $k$ wavefunctions are also important when describing surface and defect states, metal-insulator junctions and electrical transport across finite crystals and linear molecular chains.\cite{Heine, Beratan, Choi, Mavropulos}

The methods developed in Ref.~\onlinecite{Kohn59} could not be extended to higher dimensions, where the results are much more limited. The first major step here was made by Des Cloizeaux, who studied the analytic structure near real $k^\prime$s, for crystals with a center of inversion.\cite{Cloizeaux64a, Cloizeaux64b} His conclusion was that the Bloch functions and energies of an isolated simple band are analytic (and periodic) in a complex strip around the real $k^\prime$s. The restriction to crystals with center of symmetry was later removed.\cite{Nenciu83} The analytic structure
has been reconsidered in a study by Avron and Simon,\cite{Avron77} who gave answers to some important, tough qualitative questions. For example, one of their conclusion was that all isolated singularities of the band energy are algebraic branch points. The topology of the Riemann surface of the Bloch functions for finite gap potentials in two dimensions has been investigated in a series of studies by Novikov et al.\cite{Novikov99} The analytic structure has been also investigated by purely numerical methods. Since it is impossible to explore numerically the entire complex plane, numerical methods cannot provide the global structure. Even so, they can provide valuable information. For example, in the complex band calculations for Si,\cite{Schulman82} or linear molecular chains,\cite{Tosatti03, Sankey02} one can clearly see how the bands are connected, even though these studies explored only the real axis of the complex energy plane.

In this paper we consider linear molecular chains, described by a  Hamiltonian of the form
\begin{equation}
    H=-{\bf \nabla}^2+v(\rr),
\end{equation}
with $v(\rr)$ periodic with respect to one of the cartesian coordinates of $\mathbb{R}^3$, let us say $z$:
\begin{equation}
v(x,y,z+b)=v(x,y,z).
\end{equation}
The wavefunctions of $H$ are Bloch waves, $\psi_{n,\lambda}(\rr)$, with the fundamental property
\begin{eqnarray}
\psi_{n,\lambda}(x,y,z+b)&=&\lambda \psi_{n,\lambda}(x,y,z),\\
\partial_z\psi_{n,\lambda}(x,y,z+b)&=&\lambda 
\partial_z\psi_{n,\lambda}(x,y,z).\nonumber
\end{eqnarray}
We derive the generic analytic structure (the Riemann surface) of $\psi_{n,\lambda}(\rr)$ and of the corresponding energy $E_n(\lambda)$ as functions of complex $\lambda$. We shall see that they are different branches of two analytic functions, $\psi_\lambda(\rr)$ and $E(\lambda)$, with an essential singularity at $\lambda=0$ and additional branch points which, generically, are of order 3, respectively 1. We show where this branch points come from, how they move when the potential is changed and, in some cases, how to estimate their location.

Our strategy is as follows. According to Bloch theorem, we can restrict $z$ to $z\in[0,b]$ and study the following class of Hamiltonians, which will be called Bloch Hamiltonians:
\begin{equation}
    H_\lambda=-{\bf \nabla}^2+v(\rr),\ z\in[0,b],
\end{equation}
with $\lambda$ referring to the following boundary conditions:
\begin{equation}
\left\{
\begin{array}{l}
    \psi(x,y,b)=\lambda\psi(x,y,0),\\
    \partial_z\psi(x,y,b)=\lambda\partial_z\psi(x,y,0),
\end{array}
\right.
\end{equation}
which define the domain of $H_\lambda$. For $z\in[0,b]$, the eigenvectors of $H_\lambda$ and their corresponding energies coincide with $\psi_{n,\lambda}(\rr)$ and $E_n(\lambda)$. Let $\rho(H_\lambda)$ denote the resolvent set of $H_\lambda$, which is composed of those points in the complex energy plane for which $(z-H_\lambda)^{-1}$ is bounded. Now, it was long known that the Green's function, $(z-H_\lambda)^{-1}$, evaluated at some arbitrary point $z\in \rho(H_\lambda)$, is analytic of $\lambda$, for any $\lambda$ in the complex plane.\cite{Simon} In Section II, we derive the local analytic structure of the eigenvectors and eigenvalues as functions of $\lambda$, starting from this observation alone. To obtain the global analytic structure, we start with a simple $v(\rr)$, with known global analytic structure, and then study how this structure changes when $v(\rr)$ is modified.

When this formalism is applied to 1D periodic systems, all the conclusions (and a few additional ones) of Ref.~\onlinecite{Kohn59} follow with no extra effort. This is done in Section III. Section IV presents the results for linear chains. We have two applications: a compact expression for the Green's function, which is developed in Section V, and the computation of the asymptotic behavior of the density matrix for insulating linear molecular chains, which is done in Section VI. We conclude with remarks on how to calculate the analytic structure for real systems. We also have two Appendices with mathematical details.

\section{General Formalism}

We consider here, at a general and  abstract level, an analytic family, $\{H_\lambda\}_{\lambda \in \mathbb{C}}$, of closed, possibly non selfadjoint operators. The analyticity is considered in the sense of Kato,\cite{Kato} which means that, for any $z\in \rho(H_\lambda)$,  the Green's function can be expanded as
\begin{equation}\label{Kato}
    (z-H_{\lambda^\prime})^{-1}=\sum\limits_{n=0}^\infty
    (\lambda^\prime-\lambda)^n R_n(z),
\end{equation}
with the power series converging in the topology induced by the operator norm, for any $\lambda^\prime$ in a finite vicinity of $\lambda$. We have already mentioned that the Bloch Hamiltonians form an analytic family.

In this section, we discuss the analytic structure of the eigenvalues and the associated eigenvectors of $H_\lambda$, as functions of $\lambda$, based solely on the analyticity of the Green's function. For this, we need to find ways of expressing the eigenvalues and eigenvectors using only the Green's function. The major challenge will be posed by the degeneracies.

Suppose $H_{\lambda}$ has an isolated, non-degenerate eigenvalue $E_{\lambda}$, for $\lambda$ near $\lambda_0$. Then there exists a closed contour $\Gamma$ separating $E_{\lambda_0}$ from the rest of the spectrum. For $\lambda$ in a sufficiently small vicinity of $\lambda_0$, $E_\lambda$ remains the only eigenvalue inside $\Gamma$ and we can express $E_{\lambda}$ as
\begin{equation}\label{Formula1}
    E_\lambda=Tr\int_\Gamma z(z-H_\lambda)^{-1}\frac{dz}{2\pi i}.
\end{equation}
As shown in Appendix A, Eqs.~(\ref{Kato}) and (\ref{Formula1}) automatically imply that $E_\lambda$ is analytic at $\lambda_0$. Since $\lambda_0$ was chosen arbitrarily, we can conclude that the non-degenerate eigenvalues are analytic functions of $\lambda$, as long as they stay isolated from the rest of the spectrum.

\begin{figure}
  \includegraphics[width=6.0cm]{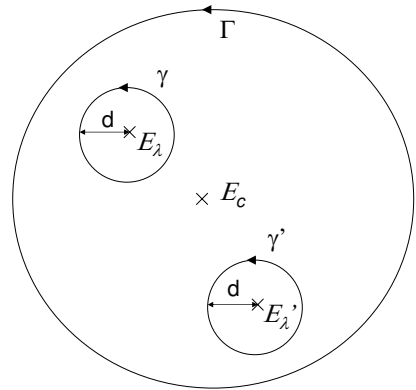}\\
  \caption{The figure shows two isolated eigenvalues of $H_\lambda$, $E_\lambda$ and $E_\lambda^\prime$, that become equal for $\lambda=\lambda_c$. The figure also shows the contours of integration, $\Gamma$, $\gamma$ and  $\gamma^\prime$ used in the text.}
\end{figure}

Suppose now that $H_\lambda$ has two isolated, non-degenerate eigenvalues, $E_\lambda$ and $E^{\prime}_\lambda$, which become equal at some $\lambda_c$ (see Fig.~1):
\begin{equation}
    E_{\lambda_c}=E^{\prime}_{\lambda_c}=E_c.
\end{equation}
We are interested in the analytic structure of these eigenvalues and the associated eigenvectors, for $\lambda$ in a vicinity of $\lambda_c$.

Eq.~(\ref{Formula1}) is no longer useful, since there is no such $\lambda$-independent contour $\Gamma$ that isolates one eigenvalue from the rest of the spectrum, for all $\lambda$ in a vicinity of $\lambda_c$. The key is to work with both eigenvalues, since we can still find a $\lambda$-independent contour $\Gamma$, separating $E_\lambda$ and $E_\lambda^\prime$ from the rest of the spectrum (see Fig.~1), as long as $\lambda$ stays in a sufficiently small vicinity of $\lambda_c$. We define,
\begin{equation}\label{theF}
    F_m(\lambda)=(E_\lambda)^m+(E^\prime_\lambda)^m,\ \ m=1,2,\ldots.
\end{equation}
The main observation is that we can use the Green's function to express $F_m(\lambda)$:
\begin{equation}\label{Formula2}
    F_m(\lambda)=Tr \int_\Gamma
    z^m(z-H_\lambda)^{-1}\frac{dz}{2\pi i}.
\end{equation}
Then, as shown in Appendix A, it follows from Eqs.~(\ref{Kato}) and (\ref{Formula2}) that $F_m(\lambda)$ are analytic functions near and at $\lambda_c$.

The functions introduced in Eq.~(\ref{theF}) provide the following representation:
\begin{eqnarray}
    E_\lambda&=&\frac{1}{2}\left[F_1(\lambda)+
    \sqrt{2F_2(\lambda)-F_1(\lambda)^2}\right]  \\
    E^{\prime}_\lambda &=& \frac{1}{2}\left[F_1(\lambda)-
    \sqrt{2F_2(\lambda)-F_1(\lambda)^2}\right].\nonumber
\end{eqnarray}
Thus, we managed to express the eigenvalues in terms of the Green's function alone. The analytic function,
\begin{equation}\label{G}
    G(\lambda)\equiv 2F_2(\lambda)-F_1(\lambda)^2,
    \label{theG}
\end{equation}
must have a zero at $\lambda_c$, so its generic behavior near
$\lambda_c$ is
\begin{equation}
    G(\lambda)=(\lambda-\lambda_c)^Kg(\lambda),
\end{equation}
with $K$ an integer larger or equal to 1 and $g(\lambda)$ analytic and non-zero at $\lambda_c$. The cases when $K>2$ are very special and will not be considered here. Only the following two possibilities are relevant to us:
\begin{equation}
    G(\lambda)=\left \{ \begin{array}{l}
                 (\lambda-\lambda_c)g(\lambda) \ \ \ \text{(type I)}\\
                 (\lambda-\lambda_c)^2g(\lambda) \ \ \text{(type II)},\\
               \end{array}
    \right .
\end{equation}
where $g(\lambda)$ is analytic and has no zeros in a vicinity of $\lambda_c$.

For a type I degeneracy, as $\lambda$ loops around $\lambda_c$, $E_\lambda$ becomes $E^\prime _\lambda$ and vice versa. The two eigenvalues are different branches of a double-valued analytic function, with a branch point of order 1 at $\lambda_c$:
\begin{equation}\label{Energy}
    E_\lambda=E_c+\alpha (\lambda-\lambda_c)^{1/2}+\ldots.
\end{equation}

For a type II degeneracy, both $E_\lambda$ and $E^\prime_\lambda$ are analytic near $\lambda_c$.

We consieder now the spectral projectors, $P_\lambda$ and $P^\prime_\lambda$, associated with $E_\lambda$ and $E^\prime_\lambda$, respectively. For $\lambda \neq \lambda_c$, they have the following representation:
\begin{equation}
    P_\lambda=\int_\gamma(z-H_\lambda)^{-1}\frac{dz}{2\pi i},
\end{equation}
where $\gamma$ is defined by $|z-E_\lambda|=d$, with $d$ small enough (thus $\lambda$ dependent) so $E^\prime_\lambda$ lies outside $\gamma$ (see Fig.~1). $P^\prime_\lambda$ has a completely equivalent representation. We list the following properties:
\begin{eqnarray}
    P_\lambda^2=P_\lambda, \ \ P_\lambda^{\prime 2}=P^\prime_\lambda \nonumber \\
    P_\lambda P_\lambda^\prime=P_\lambda^\prime P_\lambda=0.
\end{eqnarray}

For a type I degeneracy, as $\lambda$ loops around $\lambda_c$, $P_\lambda$ becomes $P_\lambda^\prime$ and vice versa. Thus, $P_\lambda$ and $P_\lambda^\prime$ are different branches of a double-valued analytic function, with branch point of order 1 at $\lambda_c$. In other words, they are given by the same function (identified from now on with $P_\lambda$), which is evaluated on different Riemann sheets. These sheets have $\lambda_c$ as a common point. Thus, if $P_\lambda$ does not diverge at $\lambda_c$, then
\begin{equation}
    \lim_{\lambda \rightarrow \lambda_c}P_\lambda=
    \lim_{\lambda \rightarrow \lambda_c}P^\prime_\lambda.
\end{equation}
But this will contradict, for example, $P_\lambda P_\lambda^\prime=0$. We must conclude that $P_\lambda$ diverges at $\lambda_c$. To find out the form of the singularity, we observe that, if $\Delta E_\lambda$ denotes the difference $E_\lambda-E_\lambda^\prime$, then $\Delta E_\lambda P_\lambda$ has a well defined limit at $\lambda_c$, which can be seen from the following representation:
\begin{equation}
    \Delta E_\lambda P_\lambda=\int_{\Gamma}(z-E_\lambda^\prime)
    (z-H_\lambda)^{-1}\frac{dz}{2\pi i}.
\end{equation}
Since $\Delta E_\lambda
\propto (\lambda-\lambda_c)^{1/2}$, we can conclude that the singularity of $P_\lambda$ is of the form $(\lambda-\lambda_c)^{-1/2}$. In addition, we mention that the total spectral projector,
\begin{equation}
    P(\lambda)=P_\lambda+P_\lambda^\prime,
\end{equation}
is analytic near and at $\lambda_c$, as it can be seen from the representation
\begin{equation}
    P(\lambda)=\int_{\Gamma}(z-H_\lambda)^{-1}\frac{dz}{2\pi i},
\end{equation}
and that the Green function for $\lambda=\lambda_c$ and $z$ near $E_c$ has the following structure:
\begin{eqnarray}
    (z-H_{\lambda_c})^{-1}=(z-E_c)^{-2}\lim_{\lambda\rightarrow\lambda_c}
    \Delta E_\lambda P_\lambda \nonumber \\
    +(z-E_c)^{-1}P(\lambda_c)+R(z),
\end{eqnarray}
with $R(z)$ analytic near $E_c$. This expression is interesting because the Green's function of self-adjoint operators, viewed as functions of $z$, always have simple poles. Thus, $H_{\lambda_c}$ cannot be a self-adjoint operator. In other words, type I degeneracies cannot occur for those values of $\lambda$ for which $H_\lambda$ is self-adjoint.

We now turn our attention to the eigenvectors. Since, for $\lambda \neq \lambda_c$, $P_\lambda$ is rank one, it can be written as
\begin{equation}
    P_\lambda=|\psi_\lambda\rangle \langle \bar{\psi}_\lambda|,
\end{equation}
with $\psi_\lambda$ ($\bar{\psi}_\lambda$) the eigenvector to the left (right), normalized as
\begin{equation}
    \langle \bar{\psi}_\lambda,\psi_\lambda \rangle=1.
\end{equation}
As explained in Ref.~\onlinecite{Nenciu91}, passing from the projector to the eigenvectors is not a trivial matter, since these vectors are defined up to a phase factor. The question is, can we choose or define these phases so that no additional singularities are introduced? We can give a positive answer when there is an anti-unitary transformation, $Q$, such that:
\begin{equation}
    H_\lambda^\dagger=QH_\lambda Q^{-1}.
\end{equation}
As we shall later see, such $Q$ exists, for example, when $v(x,y,z)=v(x,y,-z)$. Now fix an arbitrary $\psi$ and observe that
\begin{equation}\label{proj}
    P_\lambda=\frac{P_\lambda|\psi\rangle \langle
    Q\psi|P_\lambda}{\langle P_\lambda^\dagger
    Q\psi,P_\lambda\psi\rangle}.
\end{equation}
One immediate problem with the above expression is that the denominator can be zero. This can happen only for isolated values of $\lambda$ for, otherwise, the denominator will be identically zero. Let $\lambda_0$ be such value, assumed different from the branch point (we can always choose a $\psi$ satisfying this condition). We know that the only singularity of $P_\lambda$ is at $\lambda_c$, so what we have is two functions that are equal on a domain surrounding $\lambda_0$ but excluding $\lambda_0$, and one of them, $P_\lambda$, is analytic at $\lambda_0$. Then it is a fact that both functions are analytic at $\lambda_0$. This means the numerator  in the right hand side of Eq.~(\ref{proj}) must also cancel at $\lambda_0$, with the same power as the denominator and the problem disappears. 
Then, it is natural to think that we can define the left and right eigenvectors of $H_\lambda$ as:
\begin{equation}
|\psi_\lambda\rangle=\frac{P_\lambda|\psi\rangle }{\langle
P_\lambda^\dagger
    Q\psi,P_\lambda\psi\rangle^{1/2}}
\end{equation}
and
\begin{equation}
 |\bar{\psi}_\lambda\rangle=\frac{P_\lambda^\dagger Q|\psi\rangle
}{\langle P_\lambda^\dagger
    Q\psi,P_\lambda\psi\rangle^{1/2}}
\end{equation}
Using the properties of $Q$, we can equivalently write:
\begin{equation}\label{def}
|\psi_\lambda\rangle=\frac{P_\lambda|\psi\rangle }{\langle
QP_\lambda\psi,P_\lambda\psi\rangle^{1/2}}, \
|\bar{\psi}_\lambda\rangle=Q|\psi_\lambda\rangle.
\end{equation}
The only problem is the following: the denominator in Eq.~(\ref{proj}) behaves like $(\lambda-\lambda_0)^\alpha$ near $\lambda_0$, with $\alpha$ some unknown power. As we already seen, the numerator of Eq.~(\ref{proj}) must have exactly the same behavior. Now let us look at Eq.~(\ref{def}): the denominator of $\psi_\lambda$ behaves like $(\lambda-\lambda_0)^{\alpha/2}$, but the numerator can behave like $(\lambda-\lambda_0)^r$ with $r$ arbitrary, as long as the numerator of $\bar{\psi}_\lambda$ behaves like $(\lambda-\lambda_0)^{\alpha-r}$. However, we show in the following the $r$ is exactly $\alpha/2$. If $Q$ cannot be defined, then this is no longer true. You may think that, in such cases, we should modify the denominator of $\psi_\lambda$ to $\langle QP_\lambda\psi,P_\lambda\psi\rangle^r$. The problem is that $\lambda_0$ is not unique so this may fix the problem at $\lambda_0$ but not at other similar points.

Now let
\begin{equation}
    P_\lambda=\sum\limits_{i=0}^\infty (\lambda-\lambda_0)^iP_i
\end{equation}
be the expansion of $P_\lambda$ near $\lambda_0$. If the numerator of Eq.~(\ref{proj}) cancel at $\lambda_0$, an expansion in powers of $\lambda$ will show that this cancelation is equivalent to:
\begin{equation}
	P_i|\psi\rangle=0, \text{for all} \ i<K,
\end{equation}
with $K$ an integer larger or equal to 1. This also means
\begin{equation}
	P_\lambda|\psi\rangle=\sum\limits_{i=K}^\infty (\lambda-\lambda_0)^iP_i|\psi\rangle\equiv (\lambda-\lambda_0)^K|\phi_\lambda\rangle.
	\end{equation}
 Let us assume, for the beginning, that $K=1$, in which case $P_1|\psi \rangle \neq 0$. The numerator of Eq.~(\ref{proj}) becomes
\begin{equation}
	| P_\lambda\psi\rangle \langle P_\lambda \psi |Q=(\lambda-\lambda_0)^2 | P_1\psi\rangle \langle 	P_1 \psi |Q + o(\lambda-\lambda_0)^3
\end{equation}
and the denominator
\begin{equation}
	\langle Q P_\lambda \psi, P_\lambda \psi \rangle = (\lambda-\lambda_0)^2\langle Q P_1\psi,P_1 	\psi \rangle+o(\lambda-\lambda_0)^3.
\end{equation}
Although $P_1|\psi \rangle \neq 0$, we are not automatically guaranteed that $\langle Q P_1\psi,P_1 \psi \rangle \neq 0$. However, we already argued that numerator and denominator must cancel with the same power, so this must be so. We can repeat the same arguments for arbitrary $K$ and the conclusion will be the same:
\begin{eqnarray}
	\langle Q P_\lambda \psi, P_\lambda \psi \rangle = (\lambda-\lambda_0)^{2K}g(\lambda) \\
	| P_\lambda\psi\rangle=(\lambda-\lambda_0)^K|\phi_\lambda\rangle, \nonumber
\end{eqnarray}
with $g(\lambda)$ and $|\phi_\lambda\rangle$ analytic and non-zero near $\lambda_0$. Thus, we can take the square root in Eq.~(\ref{def}) without introducing a branch point. We can also see that the denominator and numerator in Eq.~(\ref{def}) cancel at $\lambda_0$ with the same power, so there is no pole at $\lambda_0$. The conclusion is that $\psi_\lambda$ is analytic at $\lambda_0$. 

There will be, inherently, a branch point at $\lambda_c$, where $\psi_\lambda$ and $\bar{\psi}_\lambda^\ast$
behave as
\begin{eqnarray}\label{vect}
    \psi_\lambda(\rr) &=& \frac{c(\rr)}{(\lambda-\lambda_c)^{1/4}}+d(\rr)+\ldots , \\
    \bar{\psi}_\lambda^\ast(\rr) &=& \frac{\bar{c}(\rr)}{(\lambda-\lambda_c)^{1/4}}+\bar{d}(\rr)+\ldots
\end{eqnarray}
i.e. $\psi_\lambda$ and $\bar{\psi}_\lambda^\ast$ have a branch point of order 3 at $\lambda_c$. If the operator $Q$ with the above mentioned properties exists, this is their only singularity near $\lambda_c$. 

For a type II degeneracy, $P_\lambda$ and $P_\lambda^\prime$ are analytic  near $\lambda_c$. The eigenvectors can be introduce in the same way as above and, if $Q$ exists, they are analytic functions of $\lambda$.

{\it Analytic deformations}. We analyze now what happens when an analytic potential $w$ is added:
\begin{equation}
    H_{\lambda,\gamma}=H_\lambda+\gamma w.
\end{equation}
By analytic potential we mean that $\{H_{\lambda,\gamma}\}_{\lambda,\gamma\in \mathbb{C}}$ is an analytic family in the sense of Kato, in both $\lambda$ and $\gamma$
(see Appendix B).

For any fixed $\gamma$, the isolated, non-degenerate eigenvalues of $H_{\lambda,\gamma}$ remain analytic of $\lambda$. The interesting question is what happens with the degeneracies. Suppose that, at some fixed $\gamma_0$, there are two isolated, non-degenerate eigenvalues, $E_{\lambda,\gamma_0}$ and $E_{\lambda,\gamma_0}^\prime$, which become equal at $\lambda_c$. For $\lambda$ in a small vicinity of $\lambda_c$ and $\gamma$ in a small vicinity of $\gamma_0$, we can define the functions $F_{1,2}(\lambda,\gamma)$ and $G(\lambda,\gamma)$ as before, which are now analytic functions in both arguments, $(\lambda,\gamma)$,
near $(\lambda_c,\gamma_0)$.

If at $\gamma_0$, $\lambda_c$ is a type I degeneracy, the only effect of a variation in $\gamma$ is a shift of $\lambda_c$. Indeed, since
\begin{equation}
    G(\lambda_c,\gamma_0)=0, \ \ \partial_\lambda G(\lambda_c,\gamma_0)\neq0,
\end{equation}
the analytic implicit function theorem assures us that there is a unique $\lambda_c(\gamma)$ such that
\begin{equation}
    G(\lambda_c(\gamma),\gamma)=0.
\end{equation}
Moreover, the zero is simple, i.e. $\lambda_c(\gamma)$ remains a type I degeneracy. For $\gamma$ near $\gamma_0$,
\begin{equation}
    \lambda_c(\gamma)=\lambda_c-\frac{\partial_\gamma
    G(\lambda_c,\gamma_0)}{\partial_\lambda
    G(\lambda_c,\gamma_0)}(\gamma-\gamma_0)
    +\ldots,
\end{equation}
where the dots indicate higher order terms in $\gamma-\gamma_0$. From Eqs.~(\ref{Energy}) and (\ref{vect}) we readily find:
\begin{equation}\label{shift}
    \lambda_c(\gamma)=\lambda_c\left[1-\frac{\gamma-\gamma_0}{\alpha}
    \int \bar{c}(x)w(x)c(x)dx+\ldots\right].
\end{equation}
If $H_{\lambda,\gamma}$ and $H_{\lambda^{\ast},\gamma}$ have complex conjugate eigenvalues, then:
\begin{equation}
    G(\lambda,\gamma)=G(\lambda^\ast,\gamma)^\ast.
\end{equation}
In this case, if $\lambda_c$ is located on the real axis, so it is $\lambda_c(\gamma)$. If not, then
\begin{equation}
    G(\lambda_c(\gamma),\gamma)=G(\lambda_c(\gamma)^\ast,\gamma)=0,
\end{equation}
which contradicts the uniqueness of $\lambda_c(\gamma)$.

Generically, a type II degeneracy splits into a pair of type I degeneracies when $\gamma$ is varied. Indeed, since
\begin{equation}
    G(\lambda_c,\gamma_0)=\partial_\lambda G(\lambda_c,\gamma_0)=\partial_\gamma
    G(\lambda_c,\gamma_0)=0,
\end{equation}
the generic structure of $G(\lambda,\gamma)$ near
$(\lambda_c,\gamma_0)$ is:
\begin{equation}
    G(\lambda,\gamma)=a(\lambda-\lambda_c)^2+b(\lambda-\lambda_c)(\gamma-\gamma_0)+c(\gamma-\gamma_0)^2+\ldots.
\end{equation}
Thus, for $\gamma\neq \gamma_0$, $G(\lambda,\gamma)$ will have, generically, two simple zeros (type I degeneracies) at:
\begin{equation}
    \lambda_c^\pm(\gamma)=\lambda_c+\frac{\gamma-\gamma_0}{2a}\left[-b \pm \sqrt{b^2-4ac}\right]+\ldots.
\end{equation}
The coefficients $a$, $b$ and $c$ can be derived from a perturbation expansion of Eqs. (\ref{theF}) and (\ref{theG}), leading to
\begin{eqnarray}\label{split}
    & &\lambda_c^\pm(\gamma)=\lambda_c-\frac{\gamma-\gamma_0}{\partial_\lambda\Delta
    E_{\lambda_c}}\times \nonumber \\
    & & \left[Tr\left\{(P_{\lambda_c}-P^\prime_{\lambda_c})w\right\} \pm
    \sqrt{-Tr \{P_{\lambda_c}w P^\prime_{\lambda_c}w\}}\right],
\end{eqnarray}
plus higher orders in $\gamma-\gamma_0$.

If a type II degeneracy does not split, it remains a type II degeneracy for all values of $\gamma$. This is a consequence of the fact that, if two analytic functions are equal on an interval, they are equal on their entire domain.

\begin{figure}
  \includegraphics[width=8.6cm]{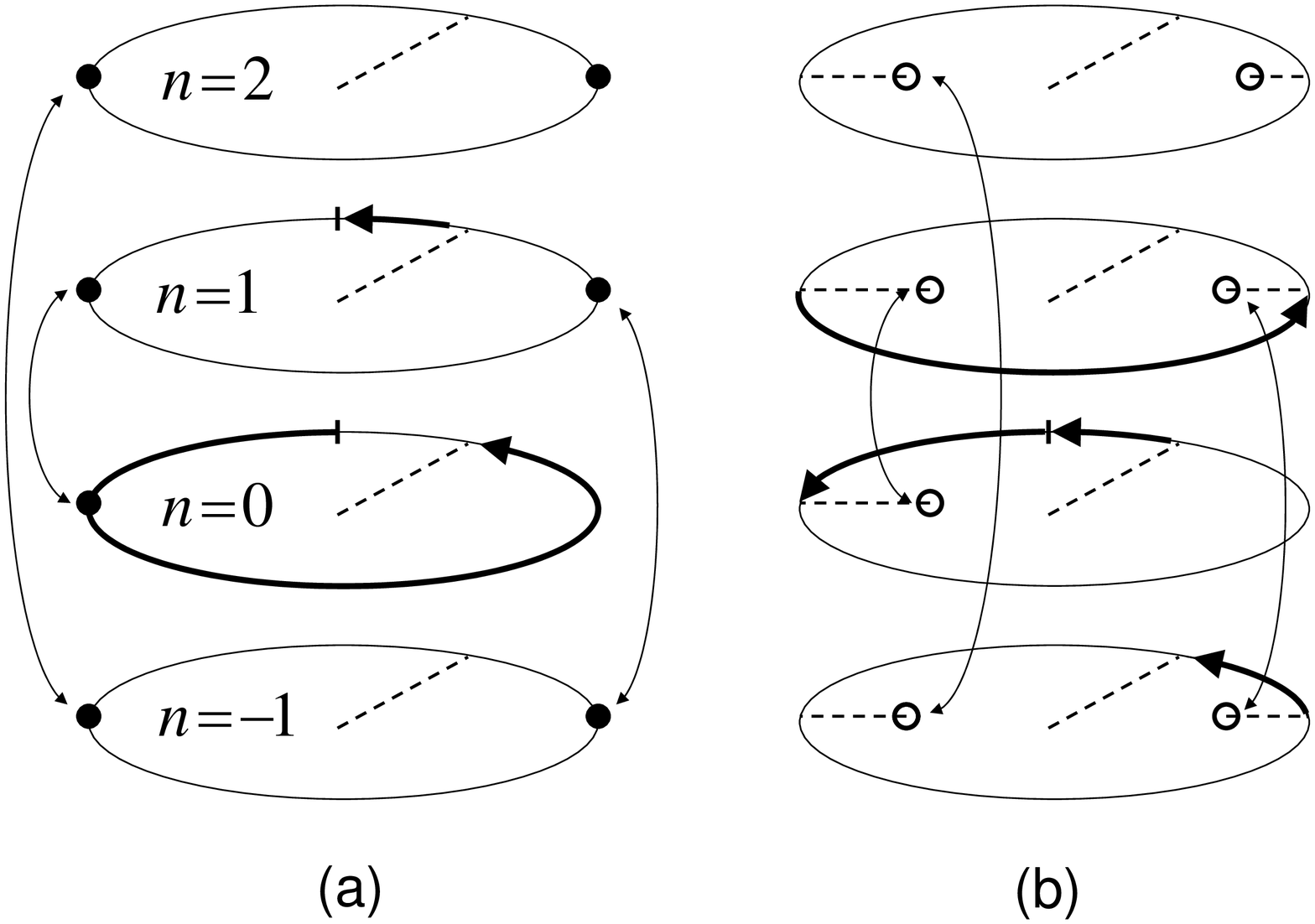}\\
  \caption{a) The Riemann surface of $E^0(\lambda)$ is a spiral, which have been cut along the dotted lines in individual sheets, indexed by $n=0,\pm1,\ldots$. The solid dots indicate the type II degeneracies and the arrows indicate how they pair. b) The Riemann surface of $E(\lambda)$ at $\gamma>0$. The empty dots represent the branch points and the arrows indicate how the Riemann sheets are connected. In both panels, the thick line shows the trajectory on the Riemann surface, when $\lambda$ moves on the unit circle.}
\end{figure}

We now summarize the findings of this Section. The isolated, non-degenerate eigenvalues of $H_\lambda$ are analytic of $\lambda$. Double degeneracies can be of different kinds. Type I degeneracies are equivalent to branch points. Near such degeneracies, $E_\lambda$ behaves as a square root. Type I degeneracies are robust to analytic perturbations: as long as they stay isolated, variations of the periodic potential cannot destroy or create but only shift them. At type II degeneracies, the eigenvalues are analytic. Type II degeneracies are unstable to analytic perturbations: generically, they split into two type I degeneracies. The other types of degeneracies were considered rare and not discussed here. The analytic structure of the spectral projectors can be automatically deduced from the analytic structure of $E(\lambda)$. If there exists an anti-unitary transformation, $Q$, such that $H^\dagger_\lambda=Q H_\lambda Q^{-1}$, then the phase of the eigenvectors can be chosen in a canonical way and their analytic structure follows automatically from the analytic structure of $E(\lambda)$. If such $Q$ does not exists, the eigenvectors will still have singularities of the type $(\lambda-\lambda_c)^{-1/4}$ at the branch points of $E_\lambda$ but the present analysis does not rule the existence of additional singularities.

A similar analysis can be developed for higher degeneracies. For a triple degeneracy, for example, we will have to consider the functions $F_m(\lambda)$, with $m=1,2,3$. However, we regard higher degeneracies as non-generic, i.e. more rare than simple degeneracies and will not be considered in this paper.

\begin{figure}
  \includegraphics[width=6cm,angle=90]{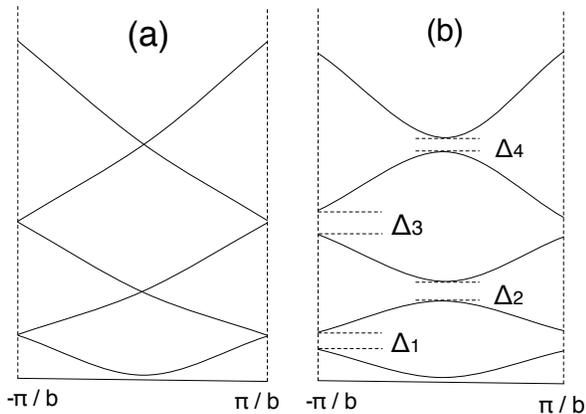}\\
  \caption{a) The plot of $E_n^0(\lambda)$ ($n=0,\pm1,\ldots$) as functions of real $k_z$ ($\lambda=e^{ik_zb}$). b) The generic band structure when the periodic potential is turned on.}
\end{figure}

\section{Strictly 1D systems}

We apply here the abstract formalism to an already well studied problem: the analytic structure of Bloch functions in 1D, i.e the wave functions of the following Hamiltonian:
\begin{equation}
    H=-\partial_x^2+v(x), \ v(x+b)=v(x),\ x\in\mathbb{R}.
\end{equation}
According to Bloch theorem, finding the wave functions and their corresponding energies is equivalent to studying the following analytic family of Hamiltonians:\cite{Simon}
\begin{equation}
    H_\lambda=-\left(\partial_x^2\right)_\lambda+v(x), \ x\in [0,b],
\end{equation}
defined in the Hilbert space of square integrable functions over the interval $[0,b]$. The index $\lambda$ refers to the boundary conditions
\begin{equation}
    \psi(b)=\lambda \psi(0),\ \psi^\prime(b)=\lambda \psi^\prime(0),
\end{equation}
which define the domain of $H_\lambda$. The energy spectrum of $H$ consists of all eigenvalues of $H_\lambda$, when $\lambda$ sweeps continuously the unit circle. If $\psi_{n,\lambda}(x)$ is the normalized eigenvector of $H_\lambda$ corresponding to the eigenvalue $E_n(\lambda)$, then $\psi_{n,\lambda}(x)$ coincides on $[0,b]$ with the Bloch wave of the same energy. If we extend these functions to the entire real axis by using
\begin{equation}
    \psi_{n,\lambda}(x+mb)=\lambda^m \psi_{n,\lambda}(x),\ x\in [0,b],
\end{equation}
they will automatically satisfy the standard normalization,
\begin{equation}
    \int^{\infty}_{-\infty}\psi_{n,1/\lambda}(x)\psi_{n,\lambda^\prime}(x)
    dx=2\pi i \lambda \delta(\lambda-\lambda^\prime).
\end{equation}

Here are a few elementary properties of $H_\lambda$. $H_\lambda$ is an analytic family in the sense of Kato, for all $\lambda\in \mathbb{C}$. $H_\lambda$ is self-adjoint if and only if $\lambda$ is on the unit circle. In general, $H_{1/\lambda^\ast}$ is the adjoint of $H_\lambda$. If $C$ is the complex conjugation, $CH_\lambda C=H_{\lambda^\ast}$. Thus, $H_\lambda$ and $H_{\lambda^\ast}$ have complex conjugated eigenvalues and $H_\lambda$ and $H_{1/\lambda}$ have identical eigenvalues. This tells us that the analytic structure is invariant to $\lambda\rightarrow 1/\lambda$ and $\lambda\rightarrow \lambda^\ast$. Because of these symmetries, it is sufficient to consider only the domain $|\lambda|\leq1$. For $\lambda$ not necessarily on the unit circle, the spectral projector on $E_{n,\lambda}$ is given by 
\begin{equation}
    P_{n,\lambda}(x,y)=\psi_{n,\lambda}(x)\psi_{n,1/\lambda}(y).
\end{equation}

We consider now the following class of Hamiltonians
\begin{equation}
    H_{\lambda,\gamma}=-\left(\partial^2_x\right)_\lambda+\gamma
    v(x),\ \ v(x+b)=v(x),
\end{equation}
and we adiabatically switch $\gamma$ from zero to one. As it is shown in Appendix B, potentials $v(x)$ with square integrable singularities\cite{footnote1} are analytic. Thus, the theory developed in the previous Section can be applied to a large class of
potentials.

\begin{figure}
  \includegraphics[width=7.0cm]{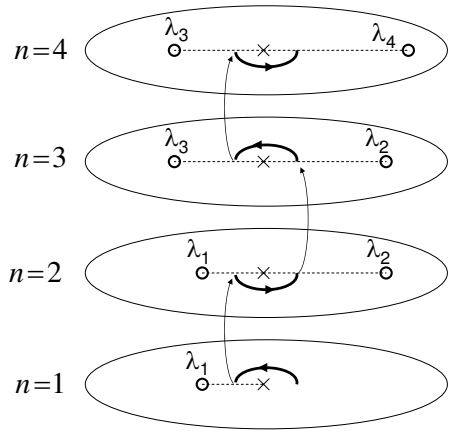}\\
  \caption{An equivalent representation of the Riemann surface of Fig.~2b. Each sheet corresponds now to one energy band. The thick line shows the trajectory on the Riemann surface when $\lambda$ encircles the origin at a small radius.}
\end{figure}

The eigenvalues and the associated eigenvectors of $H_{\lambda,0}$ are given by:
\begin{eqnarray}
    E_n^0(\lambda)&=&b^{-2}\left(2n \pi i+\ln \lambda\right)^2,
    \nonumber \\
    \psi_{n,\lambda}^0(x)&=&b^{-1/2}e^{(2n\pi i+\ln
    \lambda)x/b}.
\end{eqnarray}
They are the different branches of the following multi-valued analytic functions:
\begin{equation}
    E^0(\lambda)=b^{-2}\left(\ln \lambda\right)^2, \
    \psi_\lambda^0(x)=b^{-1/2}\lambda^{x/b}.
\end{equation}
The Riemann surface of $E^0(\lambda)$ is shown in Fig.~2a. There are only type II degeneracies and they occur at $\lambda=\pm 1$:
\begin{eqnarray}
    E_n^0(\lambda=1)&=&E_{-n}^0(\lambda=1),\nonumber \\
    E_n^0(\lambda=-1)&=&E_{-n+1}^0(\lambda=-1).
\end{eqnarray}
A plot of these eigenvalues for $\lambda$ on the unit circle is given in Fig.~3a. Now we turn on the periodic potential. Typically, the energy spectrum of $H$ consists of an infinite set of bands separated by gaps, denoted here by $\Delta_n$ (see Fig.~3b). When $v(x)$ is modified, some of the gaps may close and other may open. If we assume that all the gaps open when the periodic potential is turned on, as in Ref.~\onlinecite{Kohn59}, then all type II degeneracies split into pairs of type I degeneracies, $\lambda_c(\gamma)$ and $1/\lambda_c(\gamma)$, located on the real axis. $\lambda_c(\gamma)$ and $1/\lambda_c(\gamma)$ are branch points for $E(\lambda)$, which connect different sheets of the original Riemann surface (see Fig.~2b). Since they are constrained on the real axis, the trajectory of different branch points cannot intersect (they stay on the same Riemann sheet), i.e. the branch points remain isolated as $\gamma$ is increased. Thus, as the previous Section showed, they move analytically as we increase the coupling constant and we can conclude that the analytic structure cannot change, qualitatively, as we increase $\gamma$.

For $\gamma=0$, we move from one sheet to another as $\lambda$ moves continuously on the unit circle, as Fig.~2a shows. The situation is different for $\gamma\neq0$ (see Fig.~2b): when $\lambda$ completes one loop on the unit circle, we end up on the same point of the Riemann surface as where we started. We can then re-cut the Riemann surface, so that we stay on the same sheet when $\lambda$ moves on the unit circle (see Fig.~4).

\begin{figure}
  \includegraphics[width=8.6cm]{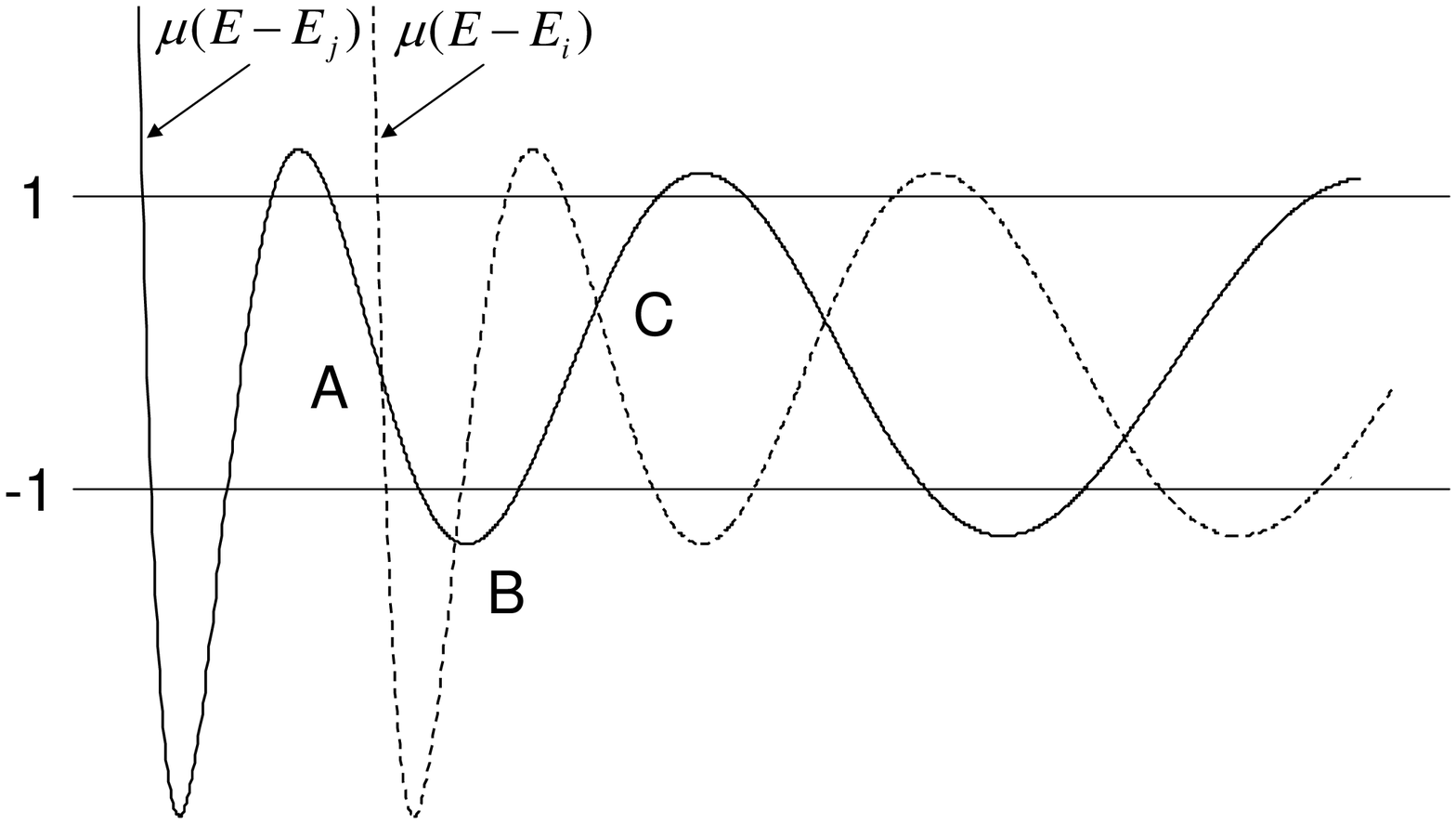}\\
  \caption{$\mu(E-E_j)$ and $\mu(E-E_i)$ as functions of $E$, for a typical Kramers function $\mu$. The solutions to Eq.~(\ref{deg2}) are given by the intersection points A, B, C, \ldots.}
\end{figure}

Thus, we rediscovered one of the main conclusions of Ref.~\onlinecite{Kohn59}. The eigenvalues $E_n(\lambda)$ are different branches of a multi-valued analytic function $E(\lambda)$ with a Riemann surface shown in Fig.~4: there are branch points or order 1 at $\lambda_1$, $\lambda_2$, $\ldots$, and an essential singularity at 0. For $\gamma$ small, Eq.~(\ref{split}) leads to
\begin{equation}
    \lambda _n=(-1)^{n-1}\left[1- \frac{b\Delta _n}{4\sqrt{\epsilon _n}}\right],
\end{equation}
where $\Delta_n$ is the $n$-th energy gap and $\epsilon_n$ is the energy in the middle of the gap.

If $v(x)=v(-x)$, we can construct $Q$ as $Q=CS$, where $C$ is the complex conjugation and $S$ is the inversion relative to $x=0$. Thus, for systems with inversion symmetry, we can also conclude at once that the only branch points of the Bloch functions are $\lambda_1$, $\lambda_2$, \ldots, which are of order 3 (see Eqs.~(\ref{vect})). The present analysis actually adds something new to the results of Ref.~\onlinecite{Kohn59}, where the author studied two particular phase choices of the Bloch functions, namely, those imposed by $\psi_\lambda(x)$ ($|\lambda|=1$) being real at the points of inversion symmetry, $x=0$ and $x=b/2$. The author warns that other choices can introduce additional singularities and thus reduce the exponential localization of the corresponding Wannier functions. A frequently used method of generating Wannier functions localized near an arbitrary $x_0$ is to impose $\text{Im} \psi_\lambda(x_0)=0$. Since such a phase choice corresponds to choosing $\psi(x)=\delta(x-x_0)$ in Eq.~(\ref{def}), we can automatically conclude that it does not introduce additional singularities and that the exponential localization of the corresponding Wannier functions is maximal.

\section{Linear molecular chains}

\begin{figure}
  \includegraphics[width=8.0cm]{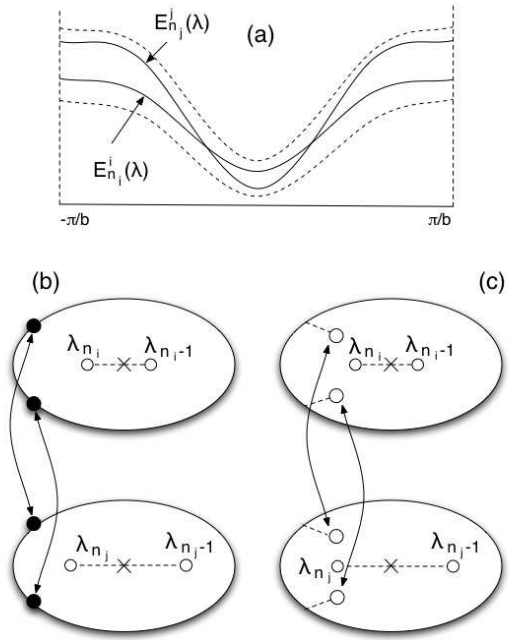}\\
  \caption{a) $E^j_{n_j}(\lambda)$ and $E^i_{n_i}(\lambda)$ as functions of real $k_z$ ($\lambda=e^{ik_zb}$) for case A. The dashed lines shows the bands after the non-separable potential was turned on. b) The Riemann sheets of $E^j_{n_j}(\lambda)$ and $E^i_{n_i}(\lambda)$ and the type II degeneracies (solid circles), with arrows indicating how they pair (case A). c) The Riemann surface at $\gamma>0$. The empty circles represent the branch points and the arrow indicate how they connect different points of the Riemann surface.}
\end{figure}

We specialize our discussion to 3D and consider Hamiltonians of the form:
\begin{equation}
    H=-{\bf \nabla}^2+V(\rr),\ \ V(x,y,z+b)=V(x,y,z).
\end{equation}
Using the Bloch theorem, we can find the spectrum and the wave functions of $H$ by studying the following family of analytic Hamiltonians:
\begin{equation}
    H_\lambda=-\partial_x^2-\partial_y^2-(\partial_z)^2_\lambda +
    V(\rr), \ z\in [0,b],
\end{equation}
with the boundary conditions
\begin{eqnarray}
    \psi(x,y,b)&=&\lambda \psi(x,y,0) \\
    \partial_z \psi(x,y,b)&=&\lambda \partial_z \psi(x,y,0).\nonumber
\end{eqnarray}
The general facts about $H_\lambda$ listed in the previous Section are still valid. Again, the analytic structure is invariant to $\lambda\rightarrow \lambda^\ast$ and $\lambda\rightarrow\ \lambda^{-1}$ so we can and shall restrict the domain of $\lambda$ to the unit disk, $|\lambda|\leq1$.

\begin{figure}
  \includegraphics[width=7.0cm]{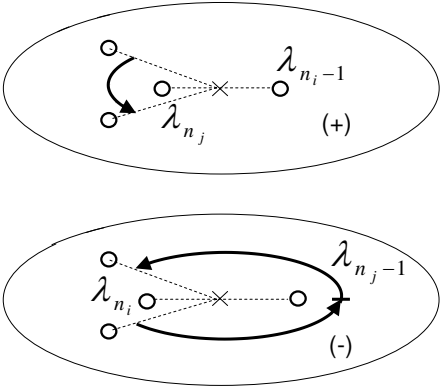}\\
  \caption{The Riemann sheets of $E_{\pm}(\lambda)$ (case A). The thick line shows a trajectory on the Riemann surface when $\lambda$ completes a loop around the origin.}
\end{figure}

We apply the analytic deformation strategy, as we did for the 1D case. We start from a Hamiltonian with known global analytic structure. For this we consider a separable potential,
\begin{equation}
    H_0=-{\bf \nabla}^2+v_\bot(x,y)+v(z),\ v(z+b)=v(z),
\end{equation}
and then adiabatically introduce the non-separable part of the Hamiltonian,
\begin{equation}
    H_\gamma=H_0+\gamma w(\rr),\ \ w(x,y,z+b)=w(x,y,z).
\end{equation}
We assume, for simplicity, that
\begin{equation}
    H_\bot\equiv -\partial_x^2-\partial_y^2+v_\bot (x,y)
\end{equation}
has only discrete, non-degenerate spectrum. For this, we will have to constrain $x$ and $y$ in a finite region, which can be arbitrarily large. To be specific, we assume $x, y\in [0,b^\prime]$ and impose periodic boundary conditions. This is actually the most widely used approach in numerical calculations involving linear chains. We also assume that $w(\rr)$ has square integrable singularities, which warranties that it is an analytic potential (see Appendix B). 

\begin{figure}
  \includegraphics[width=8.0cm]{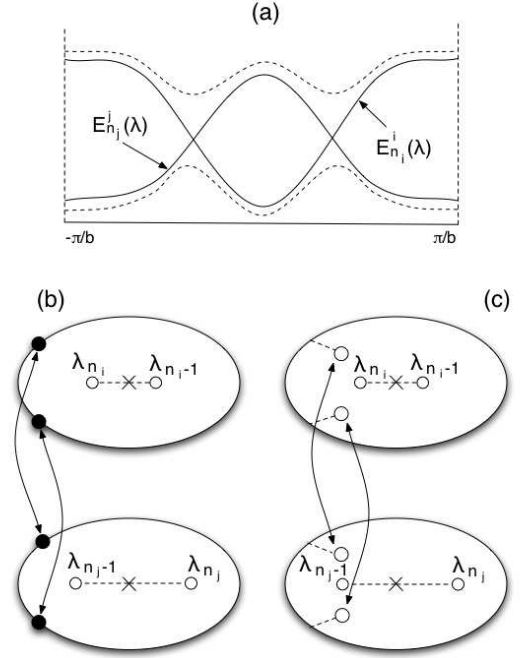}\\
  \caption{a) $E^j_{n_j}(\lambda)$ and $E^i_{n_i}(\lambda)$ as a function of real $k_z$ ($\lambda=e^{ik_zb}$) for case A. The dashed lines shows the bands after the non-separable potential was turned on. b) The Riemann sheets of $E^j_{n_j}(\lambda)$ and $E^i_{n_i}(\lambda)$ and the type II degeneracies, with arrows indicating how they pair (case C). c) The Riemann surface at $\gamma>0$. The empty circles represent the branch points and the arrow indicate how they connect different points of the Riemann surface.}
\end{figure}

Let $\phi_j(x,y)$, $E_j$ and $\psi_{n,\lambda}(z)$, $E_n(\lambda)$ denote the eigenvectors and the corresponding eigenvalues of $H_\bot$ and of
\begin{equation}
    H_{\|\lambda}\equiv-(\partial_z)^2_\lambda +v(z),
\end{equation}
respectively. Then the eigenvectors and the corresponding eigenvalues of $H_0$ are given by:
 \begin{eqnarray}
    & &\Psi^j_{n,\lambda}(x,y,z)=\phi_j(x,y)\psi_{n,\lambda}(z),\\
    & &E^j_n(\lambda)=E_j+E_n(\lambda).\nonumber 
 \end{eqnarray}
The global analytic structure of $E^j_n(\lambda)$ is known: for $j$ fixed, they are different branches of a multi-valued analytic function $E^j(\lambda)$, with a Riemann surface as in Fig.~4. We now look for type II degeneracies:
\begin{equation}\label{deg1}
    E=E_j+E_{n_j}(\lambda)=E_i+E_{n_i}(\lambda),
\end{equation}
which can occur only for $\lambda$ on the unit circle or on the real axis but away from the branch cuts. Indeed, if $\mu(E)$ denotes the Kramers function for the strictly one dimensional Hamiltonian $H_{\|\lambda}$, \cite{Kramers} then Eq.~(\ref{deg1}) is equivalent to
\begin{equation}\label{deg2}
    \mu(E-E_j)=\mu(E-E_i).
\end{equation}
In Ref.~\onlinecite{Kohn59}, it was shown that the equation $d\mu/dE=0$ has solutions only for $E$ on the real axis. Using exactly the same arguments, one can show that all the solutions of Eq.~(\ref{deg2}) are on the real axis (see Fig.~5). Given that
\begin{equation}
    \lambda^2-2\mu(E-E_j)\lambda+1=0,
\end{equation}
with $\mu(E-E_j)$ real, it follows that $\lambda$ must lie either on the unit circle or on the real axis (away from the branch cuts). For example, the solutions A and C in Fig.~5 have $\lambda$ on the unit circle, while the solution B has $\lambda$ on the real axis, inside the unit circle. We will refer to this three situations as cases A, B and C. Since the analytic structure is symmetric to $\lambda\rightarrow\lambda^\ast$, the type II degeneracies on the unit circle always come in pair, symmetric to the real axis.

\begin{figure}
  \includegraphics[width=7.0cm]{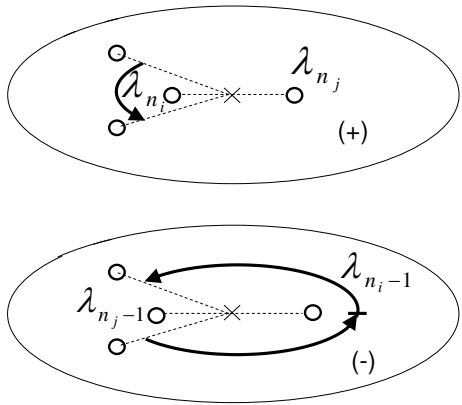}\\
  \caption{The Riemann sheets of $E_{\pm}(\lambda)$ (case C). The thick lines shows a trajectory on the Riemann surface when $\lambda$ completes a loop around the origin.}
\end{figure}

We consider first the case A, which corresponds to the case when two bands intersect as in Fig.~6a. Fig.~6b shows the Riemann sheets corresponding to these two bands. There are two type II degeneracies, marked with solid circles, on the unit circle and symmetric to the real axis. We assume for the beginning that these are the only type II degeneracies that split when the non-separable potential is turned on. When the type II degeneracies split, avoided crossings occur and the bands split (see the dashed lines in Fig.~6a). We denote the upper/lower band by $E_{\pm}(\lambda)$. When the non-separable part of the potential is turned on, the already existing branch points shift along the real axis. For small $\gamma$, the shifts can be calculated from Eq.~(\ref{shift}). In addition, two type I degeneracies appear (and another two outside the unit disk), introducing branch points that connect the original Riemann sheets. The connected sheets are shown in Fig.~6c, where we can also see that, when $\lambda$ makes a complete loop on the unit circle, we end up at the same point as where we started. This means we can re-cut the two, now connected, sheets so that we stay on the same sheet when $\lambda$ moves on the unit circle. These new sheets are shown in Fig.~7 and correspond now to the upper/lower bands $E_{\pm}(\lambda)$.

We consider now the case C, which corresponds to a situation when two bands intersect as in Fig.~8a. When the type II degeneracies split, the bands split in $E_\pm(\lambda)$ and a gap appears. This is the only qualitative difference between cases A and C. The Riemann surfaces, before and after the non-separable potential was turned on, are shown in Figs.~8b and 8c. Again, we can re-cut the Riemann surface so one sheet corresponds to one band. These new Riemann sheets are shown in Fig.~9.

The case B goes completely analogous. The qualitative differences are that the branch points split from the real axis and we don't have to re-cut the Riemann surface.

We analyze now a more involved possibility, namely when we have more type II degeneracies on the same Riemann sheet:
\begin{equation}
    E^j_{n_j}(\lambda)=E^i_{n_i}(\lambda), \ \
    E^j_{n_j}(\lambda^\prime)=E^k_{n_k}(\lambda^\prime).
\end{equation}
Such a situation appears when, for example, we have bands crossing as in Fig.~10a. The Riemann sheets for these bands and the type II degeneracies are shown in Fig.~10b. When the perturbation is turned on, the type II degeneracies split in pairs of type I degeneracies, introducing branch points connecting the original sheets as shown in Fig.~10c. Again, when $\lambda$ completes one loop on the unit circle, we end up at the same point of the Riemann surface as where we started. We can then re-cut the Riemann surface such that we stay on the same sheet when $\lambda$ moves on the unit circle (see Fig.~11). The only new element is a Riemann sheet (corresponding to the middle band) with 6 branch points.

\begin{figure}
  \includegraphics[width=8.0cm]{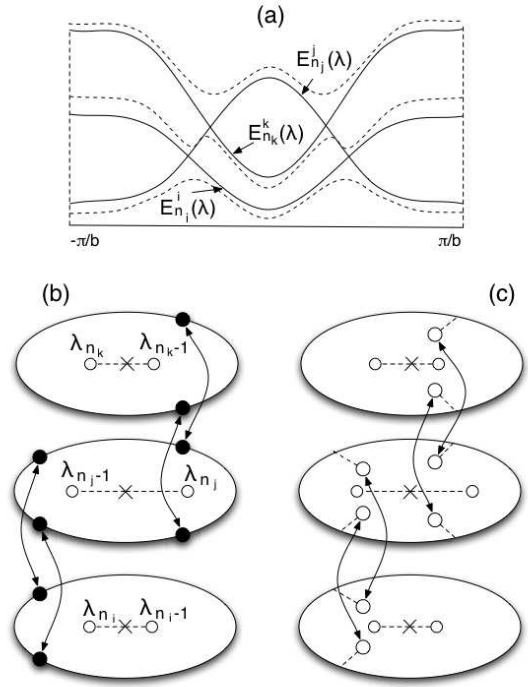}\\
  \caption{a) The band $E^j_{n_j}(\lambda)$ intersects with $E^i_{n_i}(\lambda)$ and $E^k_{n_k}(\lambda)$. The dashed lines shows the bands at $\gamma>0$. b) The Riemann sheets for these bands and the type II degeneracies, with arrows indicating how they pair. c) The Riemann sheets at $\gamma>0$.}
\end{figure}

The last situation we consider is the emergence of a complex band. Suppose that $H$ has a symmetry with an irreducible representation of dimension 2. Suppose that this symmetry is also present for the Bloch Hamiltonian at $\lambda=1$. In this case, the separable Hamiltonian will have bands that touch like in Fig.~12a. Such situations are no longer accidental. In this case, the function $G(\lambda)$ introduced in Eq.~(\ref{G}) behaves as
\begin{equation}
	G(\lambda)=(\lambda-1)^4 g(\lambda),
\end{equation}
with $g(\lambda)$ non-zero at $\lambda=1$. Following our previous notation, this will be a type IV degeneracy (see Fig.~12b). When the non-separable potential is turned on, the degeneracy at $\lambda=1$ cannot be lifted because of the symmetry. This means $G(\lambda)$ must continue to have a zero at $\lambda=1$. Generically, the order of this zero is reduced to 2 and two other zero's split, symmetric relative to the unit circle. In other words, the type IV degeneracy splits into a pair of type I degeneracies plus a type II degeneracy. $E(\lambda)$ remains analytic at $\lambda=0$, but now the two bands are entangled, in the sense that we need to loop twice on the unit circle to return back to the same point of the Riemann surface (see Fig. 12c) and we can no longer cut the Riemann surface so that we stay on the same sheet when $\lambda$ moves on the unit circle. A complex band can involve an arbitrary number of bands. For example, the new bands shown with dashed lines in Fig.~12a can entangle with other bands at $\lambda=-1$, through the same mechanism, and so on. Rather than cutting the Riemann surface in individual sheets, we think it is much more convenient to think of a complex band as living on a surface made of all the individual sheets that are entangled through the mechanism described in Fig.~12. We note that the complex band can split in simple bands as soon as the symmetry is broken. 

\begin{figure}
  \includegraphics[width=7.0cm]{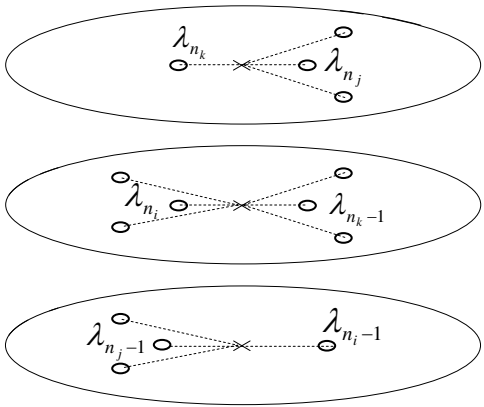}\\
  \caption{The Riemann sheets corresponding to the three
  bands (dashed lines) of Fig.~10a, for $\gamma>0$.}
\end{figure}

We can continue with further examples but we can already draw our main conclusions. The eigenvalues of $H_{\lambda,\gamma}$ are different branches of a multi-valued analytic function $E(\lambda)$. The Riemann surface of $E(\lambda)$ can be cut in sub-surfaces, such that each sub-surface describes one band. For a simple band, this subsurface consists of the entire unit disk, with cuts obtained by connecting a finite number of branch points to the essential singularity at $\lambda=0$. For a complex band, the sub-surface consists of a finite number of unit disks that are connected as in Fig.~12c.  On this sub-surface we can have an arbitrary number of branch points, that connect this sub-surface to the rest of the Riemann surface. For both simple and complex bands, the branch points are symmetric relative to the real axis and, generically, they are of order 1 (``accidental" higher degeneracies can lead to branch points of higher order).

As we analytically deform the Hamiltonian, the un-split type II degeneracies stay on the unit circle or real axis and the positions of the branch points shift smoothly. In contradistinction to the 1D case, the branch points can move from one sheet to another and their trajectories can intersect. When two of them intersect, they either become branch points of order 2 or recombine into a type II degeneracy. Higher order branch points are not stable, in the sense that small perturbations split them into two or more branch points of order 1.

The Riemann surface of the spectral projector $P_\lambda$ is the same as for $E(\lambda)$. When the inversion symmetry is present, the analytic structure of the Bloch functions can also be completely determined from the analytic structure of $E(\lambda)$: the Riemann sheets of $\psi_\lambda(x)$ are the same as for $E(\lambda)$, but the branch points are generically of order 3 (see Eq.~(\ref{vect})).

\section{The Green's function} 

With the analytic structure at hand, it is a simple exercise to find a compact expression for the Green's function $G_E\equiv (E-H)^{-1}$, which is a generalization of the well known Sturm-Liouville formula in 1D. Indeed, using the eigenfunction expansion,
\begin{equation}
G_E(\rr,\rr^\prime)=\sum_n \int_{|\lambda|=1}\frac{\psi_{n,1/\lambda}(\rr)\psi_{n,\lambda}(\rr^\prime)}{E-E_n(\lambda)}\frac{d\lambda}{2\pi i\lambda},
\end{equation}
where the sum goes over all unit disks of the Riemann surface. Changing the variable from $\lambda$ to $1/\lambda$ if necessary, the above expression can also be written as:
\begin{equation}
G_E(\rr,\rr^\prime)=\sum_n \int_{|\lambda|=1}\frac{\psi_{n,1/\lambda}(\rr_<)\psi_{n,\lambda}(\rr_>)}{E-E_n(\lambda)}\frac{d\lambda}{2\pi i\lambda},
\end{equation}
where $\rr_>=\rr$ if $z>z^\prime$, $\rr_>=\rr^\prime$ if $z^\prime>z$, and similarly for $\rr_<$. This step is necessary because we will deform the contour of integration inside the unit circle. We could deform the contour outside the unit circle, but since we chose to exclude this part of the domain we don't have this liberty anymore, and we need to re-arange the arguments before deforming the contour. The integrand (including the summation over $n$) is analytic at the branch points. Also, for $\lambda\rightarrow 0$, $E_n(\lambda)\rightarrow \infty$ and $\psi_{n,1/\lambda}(\rr_<)\psi_{n,\lambda}(\rr_>)\propto \lambda^{|z-z^\prime|}$, so there is no singularity at $\lambda=0$. Then, apart from poles, which occur whenever $E_n(\lambda)=E$, the integrand is analytic. Using the residue theorem, we conclude
\begin{equation}\label{GF}
G_E(\rr,\rr^\prime)=\sum_j\frac{\psi_{1/\lambda_j}(\rr_<)\psi_{\lambda_j}(\rr_>)}{\lambda_j \partial_\lambda E(\lambda_j)},
\end{equation}
where the sum goes over all $\lambda_j$ on the Riemann surface such that $E(\lambda_j)=E$. This expression is valid for systems with and without inversion symmetry, since it is the projector, not the individual Bloch functions, that enters into the above equations. Eq.~(\ref{GF}) is closely related to the surface adapted expression of the bulk Green's function derived in Ref.~\onlinecite{Roland}.

\section{The density matrix}

\begin{figure}
  \includegraphics[width=8.0cm]{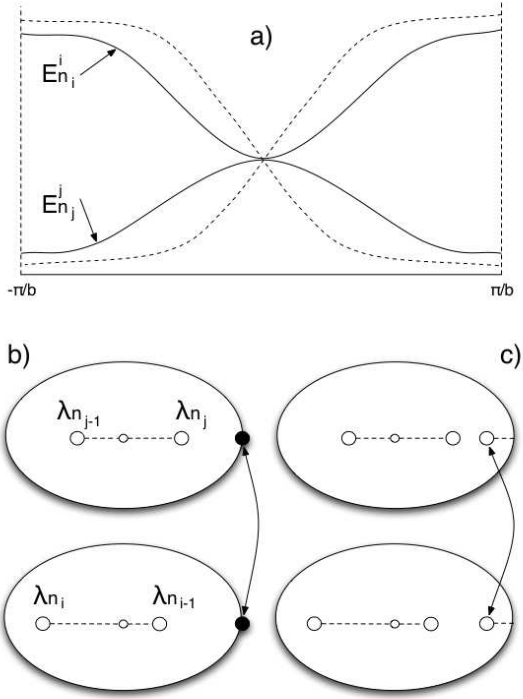}\\
  \caption{a) Two bands, $E^i_{n_i}(\lambda)$ and $E^j_{n_j}(\lambda)$, touch tangentially at $k_z=0$ ($\lambda=1$). The dashed lines show the bands after the non-separable potential was turned on. b) The Riemann sheets for these bands and the type IV degeneracy (solid circle). c) The Riemann sheets after the non-separable potential was turned on.}
\end{figure}

As a simple application, we derive the asymptotic behavior of the density matrix $n(\rr,\rr^\prime)$ for large $|z-z^\prime|$, when there is an insulating gap between the occupied and un-occupied states. We start from
\begin{equation}
n(\rr,\rr^\prime)=\frac{1}{2\pi i}\int_{\cal C}G_E(\rr,\rr^\prime)dE,
\end{equation}
where ${\cal C}$ is a contour in the complex energy plane surrounding the energies of the occupied states. Using Eq.~(\ref{GF}), we readily obtain
\begin{equation}
n(\rr,\rr^\prime)=\frac{1}{2\pi i}\int_{\gamma}\psi_{1/\lambda}(\rr_<)\psi_\lambda(\rr_>)\lambda^{-1}d\lambda,
\end{equation}
where $\gamma$ is the pre-image of the contour ${\cal C}$ on the Riemann surface of $E(\lambda)$. We now restrict $\rr$ and $\rr^\prime$ to the first unit cell and calculate the asymptotic form of $n(\rr,\rr^\prime+mb\e_z)$ for large $m$. Using the fundamental property of the Bloch functions, we have
\begin{equation}
n(\rr,\rr^\prime+mb\e_z)=\frac{1}{2\pi i}\int_{\gamma}\psi_{1/\lambda}(\rr_<)\psi_\lambda(\rr_>)\lambda^{m-1}d\lambda.
\end{equation}
We deform the contour $\gamma$ on the Riemann surface such that the distance from its points to the unit circle is maximum. In this way, we enforced the  fastest decay, with respect to $m$, of the integrand. This optimal contour, will surround (infinitely tide) the branch cuts enclosed by the original contour. The asymptotic behavior comes from the vicinity of the branch points $\lambda_c$ and $\lambda_c^\ast$ (they always come in pair) that are the closest to the unit circle. Using the behavior of the Bloch functions near the branch points, we find
\begin{equation}
n \rightarrow \text{Re}\ \bar{c}(\rr)c(\rr^\prime)\lambda_c^m \int\frac{(\lambda/\lambda_c)^{m-1} d(\lambda/\lambda_c)}{\pi(1-\lambda/\lambda_c)^{1/2}},
\end{equation}
where the integral is taken along the branch cut of $\lambda_c$. This integral is equal to $\frac{2}{\pi} B(m,1/2)$, with $B$ the Beta function. We conclude:
\begin{equation}
n(\rr,\rr^\prime+mb\e_z) \rightarrow \frac{1}{\pi}B(m,1/2)\text{Re}[\bar{c}(\rr)c(\rr^\prime)\lambda_c^m].
\end{equation}
Again, this expression holds for systems with and without inversion symmetry, since it is the projector, not the individual Bloch functions, that enters in the above equations.

\section{Conclusions}
First, we want to point out that the formalism presented here can be also applied to cubic crystals, to derive the analytic structure of the Bloch functions with respect to $k_z$, while keeping $k_x$ and $k_y$ fixed. Preliminary results and several applications have been already reported in Ref.~\onlinecite{ProdanKohn}.

We come now to the question of how to locate the branch points for a real system. In a straightforward approach, one will have to locate those $\lambda$ inside the unit disk where $H_\lambda$ displays degeneracies. Although such a program can be, at least in principle, carried out numerically, there are few chances of success without clues of where these points are located. This is because, in more than one dimension, these degeneracies occur, in general, at complex energies. One possible solution is to
follow the lines presented in this paper: locate the type II degeneracies for a separable potential $v_s$, chosen as close to the real potential $v$ as possible, and follow the trajectory of the branch points as the potential is adiabatically changed $v_\gamma=v_s+\gamma(v-v_s)$, from $\gamma=0$ to 1. We plan to complete such a program in the near future.

The analytic structure of the band energies and Bloch functions of 3D crystals, viewed as functions of several variables $k_x$, $k_y$ and $k_z$ is a much more complex problem, with qualitatively new aspects. It will be interesting to see if this problem can be tackled by the same analytic deformation technique.

The main part of this work was completed while the author was visiting Department of Physics at UC Santa Barbara. This work was part of the ``Nearsightedness" project, initiated and supervised by Prof. Walter Kohn and was supported by Grants No.~NSF-DMR03-13980,
NSF-DMR04-27188 and DOE-DE-FG02-04ER46130. The complex bands were investigated while the author was a fellow of the Princeton Center for Complex Materials. 

\appendix

\section{}

We prove here that if $\{H_\lambda\}_{\lambda\in\mathbb{C}}$ is an
analytic family in the sense of Kato,\cite{Kato} then $F_m(\lambda)$
defined in Eq. (\ref{theF}) are analytic functions. If
$R_z(\lambda)\equiv(z-H_\lambda)^{-1}$, by definition,\cite{Kato}
the limit
\begin{equation}
    \lim\limits_{\lambda^\prime \rightarrow\lambda}
    \frac{R_z(\lambda)-R_z(\lambda^\prime)}{\lambda-\lambda^\prime}
\end{equation}
exist in the topology induced by the operator norm, for any $\lambda \in \mathbb{C}$ and
$z\in\rho(H_\lambda)$. We denote this limit by $\partial_\lambda
R_z(\lambda)$. Consider now an arbitrary $\lambda_0 \in \mathbb{C}$,
and a contour $\Gamma $ in $\rho(H_{\lambda_0})$ surrounding $N$
eigenvalues of $H_{\lambda_0}$. Note that the Bloch Hamiltonians considered in this paper have compact resolvent so their spectrum is always discrete. For $\lambda$ in a small
neighborhood of $\lambda_0$, $\Gamma$ remains in $\rho(H_{\lambda})$ and we
can define
\begin{equation}
    \hat{F}_m(\lambda)\equiv\int_\Gamma z^m R_z(\lambda)\frac{dz}{2\pi
    i},
\end{equation}
and $F_m(\lambda)=Tr\hat{F}_m(\lambda)$. $\hat{F}_m(\lambda)$ is an
analytic family of rank $N$ operators for $\lambda$ in a small
neighborhood of $\lambda_0$. Indeed, if
\begin{equation}
    \hat{F}^\prime_m(\lambda)\equiv\int_\Gamma z^m \partial_\lambda R_z(\lambda)\frac{dz}{2\pi
    i},
\end{equation}
then
\begin{equation}\label{inter2}
    \left \|\frac{\hat{F}_m(\lambda)-\hat{F}_m(\lambda^\prime)}
    {\lambda-\lambda^\prime}-\hat{F}^\prime_m(\lambda)\right\|\rightarrow
    0
\end{equation}
as $\lambda^\prime\rightarrow\lambda$, since it can be bounded by
\begin{eqnarray}
    \int_\Gamma |z|^m \left\|\frac{R_z(\lambda)-R_z(\lambda^\prime)}
    {\lambda-\lambda^\prime}-\partial_\lambda
    R_z(\lambda)\right\|\frac{|dz|}{2\pi}.
\end{eqnarray}
This means the limit
\begin{equation}
    \lim\limits_{\lambda^\prime \rightarrow\lambda}
    \frac{\hat{F}_m(\lambda)-\hat{F}_m(\lambda^\prime)}
    {\lambda-\lambda^\prime}
\end{equation}
exists and is equal to $\hat{F}^\prime_m(\lambda)$. Since
$\hat{F}^\prime_m(\lambda)$ is the difference of rank $N$ operators,
it is at most rank $2N$. In particular,
$|Tr\hat{F}^\prime_m(\lambda)|<\infty$. Then
\begin{equation}\label{inter1}
    \left|\frac{F_m(\lambda)-F_m(\lambda^\prime)}
    {\lambda-\lambda^\prime}-Tr\hat{F}^\prime_m(\lambda)\right|
    \rightarrow 0
\end{equation}
as $\lambda^\prime\rightarrow\lambda$, since
\begin{eqnarray}
    \left|Tr \left[\frac{\hat{F}_m(\lambda)-\hat{F}_m(\lambda^\prime)}
    {\lambda-\lambda^\prime}-\hat{F}^\prime_m(\lambda)\right]\right|\nonumber
    \\
    \leq 4N\left \|\frac{\hat{F}_m(\lambda)-\hat{F}_m(\lambda^\prime)}
    {\lambda-\lambda^\prime}-\hat{F}^\prime_m(\lambda)\right\|,
\end{eqnarray}
and Eq.~(\ref{inter1}) follows from Eq.~(\ref{inter2}). Thus, the
limit
\begin{equation}
    \lim\limits_{\lambda^\prime \rightarrow\lambda}
    \frac{F_m(\lambda)-F_m(\lambda^\prime)}
    {\lambda-\lambda^\prime}
\end{equation}
exists and is equal to $Tr\hat{F}^\prime_m(\lambda)$.

\section{}

We discuss here the analytic perturbations for linear molecular chains. As we did in the main text, we constrain $x$ and $y$ in finite intervals. The Bloch functions are determined by the following Hamiltonian:
\begin{equation}
    H_{\lambda,\gamma}=-\Delta_\lambda+\gamma w, \ \ x,y\in [0,b^\prime] \ \text {and} \ z\in[0,b],
\end{equation}
where $\Delta_\lambda$ is the Laplace operator with periodic boundaries in $x$ and $y$ and the usual Bloch conditions in $z$. We show that if
\begin{equation}\label{condition}
    \|w\|_{L^2}\equiv \left[ \int w({\bf r})^2d{\bf r}\right]^{1/2}<\infty,
\end{equation}
then $H_{\lambda,\gamma}$ is an analytic family for all $\gamma \in \mathbb{C}$. For this we need the following technical result.

{\it Proposition.} Suppose $w$ satisfies Eq.~(\ref{condition}). Then, for $a$ positive and sufficiently large, there exists a positive $\epsilon_a$ such that $\lim \limits_{a\rightarrow\infty}\epsilon_a=0$ and:
\begin{equation}\label{goal}
    \|wf\|_{L^2}\leq \epsilon_a \|(H_{\lambda,0}+a)f\|_{L^2},
\end{equation}
for any $f$ in the domain of $H_{\lambda,0}$.

Now, pick an arbitrary $\gamma_0$, let $z \in \rho
(H_{\lambda,\gamma_0})$ and denote
$d_z=\|(H_{\lambda,\gamma_0}-z)^{-1}\|<\infty$, where $\| \ \|$
denotes the operator norm. Since
\begin{eqnarray}
    (1-\epsilon_a |\gamma_0|) \|wf\|_{L^2}\leq \epsilon_a \|(H_{\lambda,\gamma_0}+a)f\|_{L^2},
\end{eqnarray}
for any $f$ in the domain of $H_{\lambda,0}$, taking $a$
sufficiently large so that $1-\epsilon_a|\gamma_0|>0$, we obtain:
\begin{equation}
    \|w(H_{\lambda,\gamma_0}-z)^{-1}\|\leq
    \frac{\epsilon_a[1+|z+a|
    d_z]}{1-\epsilon_a |\gamma_0|}<\infty.
\end{equation}
If $M$ denotes the right hand side of the above equation, then
$(H_{\lambda,\gamma}-z)^{-1}$ is bounded for
$|\gamma-\gamma_0|<M^{-1}$ and has the following norm convergent
expansion:
\begin{eqnarray}
    (H_{\lambda,\gamma}-z)^{-1}=(H_{\lambda,\gamma_0}-z)^{-1}\nonumber
    \\
    \times \sum
    \limits _{n=0}^{\infty}(\gamma_0-\gamma)^n
    [w(H_{\lambda,\gamma_0}-z)^{-1}]^n.
\end{eqnarray}
Thus, $(H_{\lambda,\gamma}-z)^{-1}$ is analytic at the arbitrarily
chosen $\gamma_0$.

We now give the proof of the proposition. For $f$ in the domain of $H_{\lambda,0}$, let $g=(H_{\lambda,0}+a)f$. If $G_\lambda=(H_{\lambda,0}+a)^{-1}$, with $a$ assumed sufficiently large so that $(H_{\lambda,0}+a)^{-1}$ exists, we have
\begin{equation}
    f({\bf r})=\int
    G_\lambda({\bf r},{\bf r}^\prime;a)g({\bf r}^\prime)d{\bf r}^\prime
\end{equation}
and Schwartz inequality gives ($\|f\|_{L^\infty}\equiv\sup\limits_{{\bf r}}|f({\bf r})|$)
\begin{equation}\label{basic}
    \|f\|_{L^\infty}\leq \sup_{{\bf r}} \left[\int
    |G_\lambda({\bf r},{\bf r}^\prime;a)|^2 d{\bf r}^\prime
    \right]^{1/2}
    \|g\|_{L^2}.
\end{equation}
If we denote
\begin{equation}\label{alpha}
    \alpha_a=\sup_{{\bf r}}\left[\int
    |G_\lambda({\bf r},{\bf r}^\prime;a)|^2 d{\bf r}^\prime
    \right]^{1/2},
\end{equation}
with the aid of Eq.~(\ref{basic}), we obtain:
\begin{equation}\label{final}
    \|wf\|_{L^2}\leq\alpha_a\|w\|_{L^2}\|(H_{\lambda,0}+a)f\|_{L^2},
\end{equation}
i.e. Eq.~(\ref{goal}), if we identify $\epsilon_a\equiv\alpha_a\|w\|_{L^2}$. We remark that $\alpha_a$ defined in Eq. (\ref{alpha}) is optimal, in the sense that there are $f^\prime$s when we do have equality in Eq.~(\ref{final}). It remains to show that $\lim_{a\rightarrow\infty}\alpha_a=0$.

If $G^0=(-\Delta+a)^{-1}$, with $\Delta$ the Laplace operator over the entire $\mathbb{R}^3$, then we have the following representation:
\begin{equation}
    G_\lambda({\bf r},{\bf r}^\prime;a)=\sum\limits_{{\bf R}\in
    \Gamma} \lambda^{-n_z}G^0({\bf r}-{\bf r}^\prime+{\bf R};a),
\end{equation}
where the sum goes over all points of the lattice $\Gamma$ defined by ${\bf R}=(n_x b^\prime,n_y b^\prime,n_z b)$. Since $G^0$ is real and positive, we can readily see that $|G_\lambda({\bf r},{\bf r}^\prime;a)|\leq G_{|\lambda|}({\bf r},{\bf r}^\prime;a)$. Moreover,
\begin{equation}
    \int
    G_{|\lambda|}({\bf r},{\bf r}^\prime;a)^2 d{\bf
    r}^\prime=(H_{|\lambda|,0}+a)^{-2}({\bf r},{\bf r}),
\end{equation}
and we have the following representation:
\begin{equation}
    (H_{|\lambda|,0}+a)^{-2}({\bf r},{\bf r})=\sum\limits_{{\bf R}\in
    \Gamma} |\lambda|^{-n_z}C^0({\bf R};a),
\end{equation}
where $C^0=(-\Delta+a)^{-2}$. Note that the diagonal part of  $(H_{|\lambda|,0}+a)^{-2}$ is independent of ${\bf r}$. $C^0$ can be explicitly calculated, leading to:
\begin{equation}
    \alpha_a\leq \left[ \sum\limits_{{\bf R}\in
    \Gamma} |\lambda|^{-n_z}\frac{e^{-\sqrt{a}|{\bf R}|}}{2\pi
    \sqrt{a}}\right]^{1/2},
\end{equation}
with equality for $\lambda$ real and positive. The right hand side
is finite for $a$ sufficiently large and goes to zero as
$a\rightarrow \infty$.


\begin{thebibliography}{99}
\bibitem{footnote1} By a square integrable singularity we mean
$\int v(x)^2 dx<\infty$, with the integral taken over a finite
vicinity of the singularity.
\end{thebibliography}
\end{document}